# Olympic Gels: Concatenation and Swelling

*Michael Lang,[1]\* Jakob Fischer, [1,2] Marco Werner, [1] Jens-Uwe Sommer [1]*

[1]Leibniz Institut für Polymerforschung Dresden, Hohe Straße 6, 01069 Dresden, Germany. [2]now at: Bio Systems Analysis Group, Department of Computer Science, Friedrich Schiller University Jena, Ernst-Abbe-Platz 2, 07743 Jena.
E-mail: lang@ipfdd.de

**Summary:** Concatenation and equilibrium swelling of Olympic gels, which are composed of entangled cyclic polymers, is studied by Monte Carlo Simulations. The average number of concatenated molecules per cyclic polymers, $f_n$, is found to depend on the degree of polymerization, $N$, and polymer volume fraction at network preparation, $\varphi_0$, as $f_n \propto \varphi_0^{\nu/(3\nu-1)} N$ with scaling exponent $\nu = 0.588$. In contrast to chemically cross-linked polymer networks, we observe that Olympic gels made of longer cyclic chains exhibit a smaller equilibrium swelling degree, $Q \propto N^{-0.28} \varphi_0^{-0.72}$, at the same polymer volume fraction $\varphi_0$. This observation is explained by a des-interspersion process of overlapping non-concatenated rings upon swelling, which is tested directly by analyzing the change in overlap of the molecules upon swelling.

## Introduction

Olympic gels[1-3] are networks made of cyclic polymers ("rings") connected by the mutual topological inclusion of polymer strands, see Figure 1, with their elastic properties depending exclusively on the entanglements that were fixed by the linking of the rings. This particular difference to conventional polymer networks and gels makes these materials an interesting model system, since the pristine effect of entanglements on thermodynamic properties of polymers might be accessible. In particular, such gels could reveal the role of entanglements for equilibrium swelling of polymer networks, which is an outstanding problem in polymer physics[4].

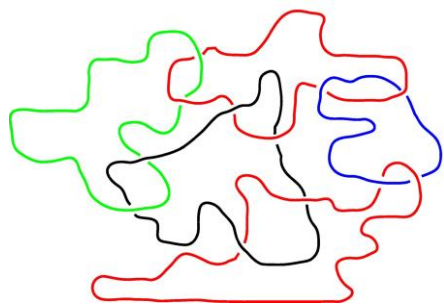

Figure 1. Sketch of an Olympic gel

Olympic gels also serve as an ideal model system to understand the threading of cyclic molecules by other polymers, which impacts the dynamics of cyclic polymers upon addition of small amounts of added linear chains[5] or is necessary to obtain a quantitative prediction for the trapping of cyclic molecules in networks[6]. The absence of concatenation between pairs of overlapping cyclic molecules leads to an additional topological repulsion between these molecules that is in the focus of recent research[7-9]. This topological repulsion is considered to be the driving force to understand the mutual compression of non-concatenated cyclic molecules in melt[10]. The mutual trapping of cyclic molecules, on the other hand, determines the structure of kinetoplast DNA[11], which poses interesting questions on its replication[12]. While the special two-dimensional arrangement of stiff kinetoplast DNA has been studied with computer simulations[13-15], our recent work[16] was first to analyze structure and properties of Olympic gels made of mono-disperse solutions of flexible polymers. The main results of this work are discussed below in a broader context as in the original letter along with concatenation data of melts of interpenetrating rings[10] and the consequences for our understanding of entanglement effects on the swelling of conventional polymer networks.

## Computer Simulations and Analysis

To simulate Olympic gels and concatenation of cyclic molecules in melt we used a GPU-version[17] of the bond fluctuation method[18], which is an efficient simulation method for polymers in the semi-dilute and concentrated regime[19]. Concatenation and swelling of Olympic gels was studied as function of the degree of polymerization N and polymer volume fraction $\varphi_0$. Two sets of samples were prepared, one with[10] and one without periodic boundaries[16] in order to study concatenation at no boundary effects and equilibrium swelling by swelling in a larger compartment respectively. In both cases, we allowed the crossing of bonds at preparation state by including additional diagonal moves without a change in the excluded volume constraints. By returning to the original set of moves after preparation, the thus created randomly concatenated topology is conserved.

After preparation, the samples with non-periodic boundaries were placed into the middle of a large simulation container and swollen to equilibrium, which was monitored by the drop in the polymer volume fraction near the middle of the gel. Empty lattice sites model a perfect athermal solvent. The equilibrium degree of swelling Q was determined by analyzing $\varphi$ for the innermost 50% of the monomers. We considered $\varphi = 0.5$ as melt concentration with reference value Q = 1. The overlap number of a given ring, P, was determined by counting the centers of mass of other cyclic molecules in a sphere with radius D around the center of mass of each ring, whereby D is the root mean square distance of two opposite monomers of a ring. For convenience, we also used D to measure the deformation of the rings. The preparation state is denoted by a subscript "0" to distinguish from swelling equilibrium. Note that the rings are only weakly deformed in the swollen state with a maximum $D/D_0 \approx 2.07$ for all samples.

Concatenation of the cyclic molecules was analyzed for all samples using tools from knot theory as described in Refs.[10,20]. The key parameters of the samples without periodic boundaries[16] prepared for swelling are summarized in Table 1. The simulation series for the concatenation study[10] contains additional samples as compared to Table 1 at similar or lower degrees of polymerization or volume fractions, see Ref. [10]. The concatenation data is summarized in Figures 2 and 3.

| Sample | #1 | #2 | #3 | #4 | #5 | #6 | #7 | #8 | #9 |
|---|---|---|---|---|---|---|---|---|---|
| N | 128 | 256 | 256 | 512 | 512 | 512 | 1024 | 1024 | 1024 |
| M | 1024 | 512 | 2048 | 2048 | 1024 | 512 | 1024 | 512 | 1024 |
| $\varphi_0$ | 0.5 | 0.5 | 0.25 | 0.5 | 0.25 | 0.125 | 0.5 | 0.25 | 0.0625 |
| $f_n$ | 2.7 | 5.64 | 2.89 | 10.9 | 6.04 | 2.98 | 17.7 | 10.1 | 2.76 |
| Q | 14.9 | 10.5 | 25.7 | 9.54 | 17.0 | 38.3 | 8.6 | 14.5 | 39.8 |
| $D_0^2$ | 339 | 632 | 857 | 1340 | 1710 | 2113 | 2505 | 3163 | 4884 |
| $D^2$ | 1027 | 2382 | 2281 | 5236 | 5452 | 5216 | 10708 | 11579 | 10621 |
| $P_0$ | 9.03 | 13.1 | 9.59 | 18.8 | 12.9 | 8.19 | 27.5 | 19.1 | 9.37 |
| P | 3.64 | 8.37 | 3.61 | 17.0 | 9.49 | 3.85 | 26.3 | 16.5 | 4.62 |

Table 1. N is the degree of polymerization of the rings, M the number of rings per sample, $\varphi_0$ the polymer volume fraction at preparation conditions, $f_n$ the average number of concatenated pairs of rings per ring, Q the equilibrium degree of swelling, $D^2$ and $D_0^2$ are the mean squared distances of two opposite monomers of a ring in the swollen and the preparation state, P and $P_0$ are the overlap numbers in the swollen state and the preparation state.

## Concatenation

For the average number of pairwise concatenations per cyclic molecule, $f_n$, we derived[10] an expression based upon the assumption that a line density of polymer strands proportional to the polymer volume fraction $\varphi_0$ at preparation conditions can concatenate

with a given cyclic molecule. Since these lines must pass at least once through an effective

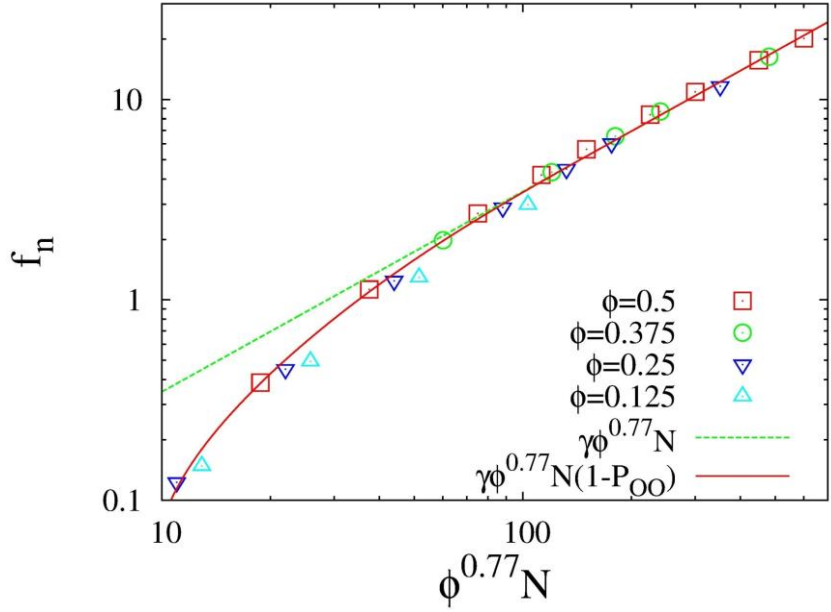

Figure 2. Average number of pairwise concatenations per ring.

area proportional to the square size of a molecule, $R^2$, we estimated for mono-disperse solutions that

$$f_n \propto \varphi_0 R^2 \approx \varphi_0^{\nu/(3\nu-1)} N \approx \varphi_0^{0.77} N, \qquad (1)$$

for $f_n > 2$. This rough estimate is well supported by the simulation data of Ref.[10], see Figure 2. The corrections for small $f_n$ arise due to a partial blocking of the penetrable area by the excluded volume of the effective monomers of the ring[21]. This idea is supported by experimental data on the trapping of small cyclic molecules in a network[22,23].

As discussed in Ref.[10], the average number of concatenations grows quicker with increasing molecular weight than the overlap number of cyclic molecules in semi-dilute solution,

$$P \propto \varphi_0 R^3/N \approx \varphi_0^{1-3(\nu-1/2)/(3\nu-1)} N^{1/2}. \qquad (2)$$

Thus, it is expected that the observed regime is the limiting case for low overlap number, while at large $P$ a second regime with $f_n \propto P$ is expected to follow. This latter regime, however, has not been observed within the range of parameters used in Ref.[10].

The probability $P_{OO}$ that a given ring is not concatenated by any other ring is shown in Figure 3. It was found that $P_{OO}$ decays exponentially as function of the average number of concatenations, $f_n \propto \varphi_0^{0.77} N$. Note that an exponential dependence can be derived from assuming a constant probability for concatenation per unit area bounded by the ring.[21] For large cyclic molecules, the data of Clarson *et al.*[22] does not follow an exponential dependence. Instead, a significantly larger portion of trapped cyclic molecules is determined. This discrepancy is possibly explained by the entropic trapping[24] of large molecules in a swollen network of heterogeneous density.

In the first theoretical estimate for concatenation it was assumed[25] that the overlap of the segment distributions in space of two possibly concatenating molecules could be used to estimate the concatenation probability. But recent work[7] demonstrates that the topological contribution is clearly different from the excluded volume contribution to repulsion and thus, different to the overlap of the segment distributions. Field theoretical models[26-28] predict a scaling of the number of concatenations $\propto N^{1/2}$ in the limit of large N, which agrees with our expectation for large N but does not explain the scaling of our data.

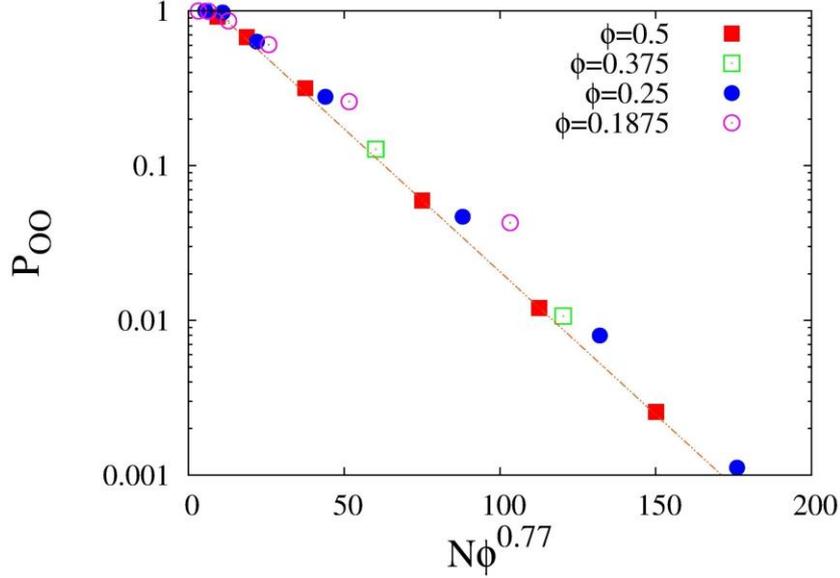

Figure 3. Probability $P_{OO}$ that a ring is not linked to any other ring decays exponentially as function of $f_n \propto \varphi_0^{0.77} N$.

## Swelling

To investigate equilibrium swelling of Olympic gels, we focus on samples with an average number of concatenations per ring $f_n \geq 2$ for which we can identify a dominant largest cluster (gel), see Table 1. In Figure 4, we display our simulation results[16] for the equilibrium swelling of Olympic gels. In contrast to the Flory-Rehner (FR) prediction[29,30] for the equilibrium degree of swelling,

$$Q = \frac{1}{\varphi} \approx N^{\frac{3(3\nu-1)}{4}} \varphi_0^{-\frac{1}{4}} \approx N^{0.57} \varphi_0^{-0.25}, \tag{3}$$

we observe a reduction of the degree of swelling with increasing chain length, see inset of

Figure 4. The best overlap of all simulation data is consistent with an ad hoc scaling law given by

$$Q \approx N^{-0.28}\varphi_0^{-0.72} \qquad (4)$$

as shown in the main plot of Figure 4. To explain this unexpected behavior requires a more detailed analysis of the swelling process that we present below.

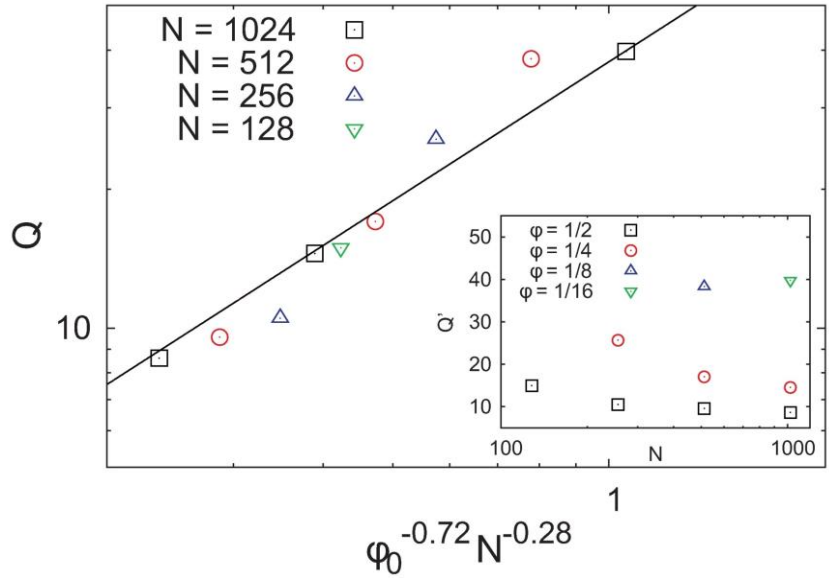

Figure 4. The scaling of the equilibrium degree of swelling. Inset: unscaled data. The line indicates the proposed scaling $Q \propto N^{-0.28}\varphi_0^{-0.72}$.

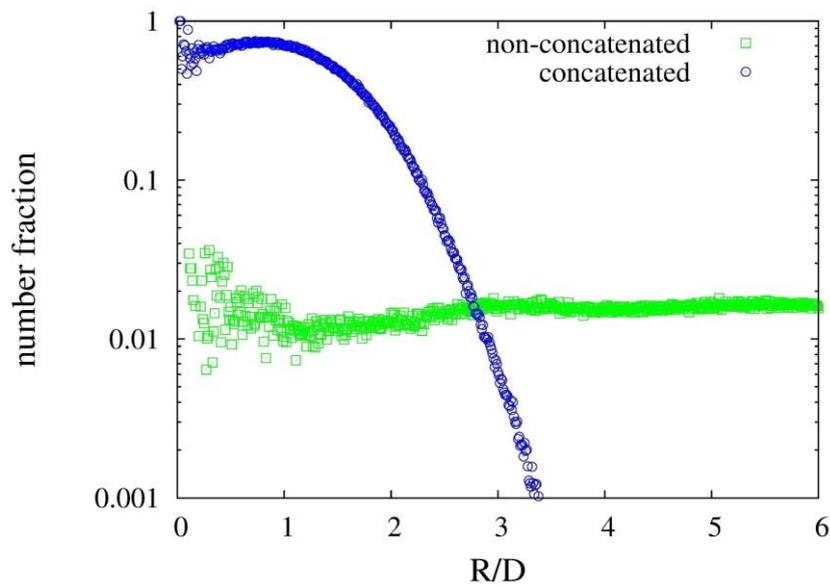

Figure 5. Normalized distance distribution between centers of mass of previously overlapping non-concatenated and concatenated rings at swelling equilibrium (sample #6).

The starting point of our analysis is the fact that the network is held together by mutual concatenations, but not all overlapping cyclic molecules are concatenated. Thus, we can distinguish between concatenated and non-concatenated rings which are overlapping at, for instance, preparation conditions. The distance distribution (distance of centers of mass) of these two ring populations is then analyzed again at swelling equilibrium. The data of the two initially overlapping populations of sample #6, see Table 1, with small $f_n$ is shown in Figure 5 as an example. The data show that non-concatenated rings essentially are squeezed out of the volume $4\pi D^3/3$ while the concatenated rings remain within a distance of order D. This tendency remains intact for all samples of our study, whereby squeezing becomes less pronounced for large $f_n$.

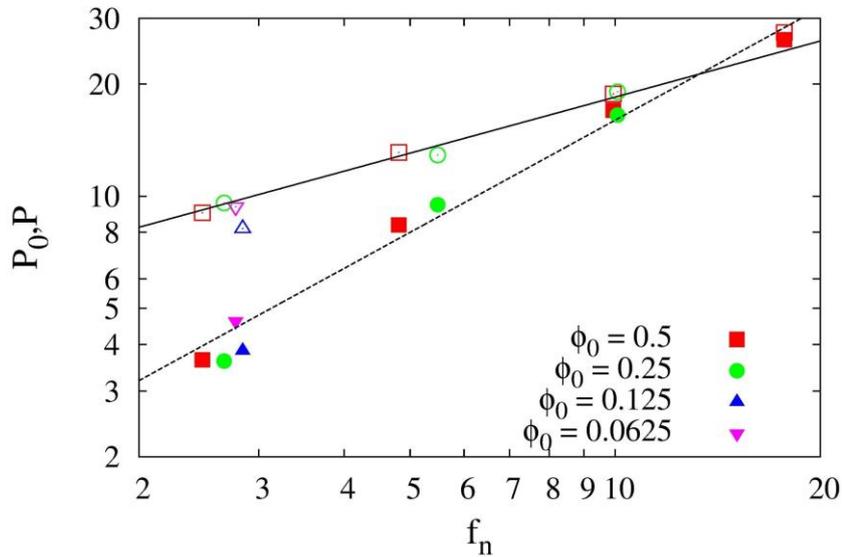

Figure 6. The overlap number of cyclic polymers in the preparation state, $P_0$ (hollow symbols), at swelling equilibrium, P (filled symbols) as function of the average number of concatenations per ring, $f_n$. The dashed line indicates $P \propto f_n$ and the solid line indicates $P_0 \propto f_n^{1/2}$.

To test this general trend, we propose a rather crude quantitative approximation: the number of overlapping molecules at swelling equilibrium, P, is roughly proportional to $f_n$. This approximation seems to work for the samples of our study as shown in Figure 6. Remember, that $f_n$ grows linearly with N and thus, quicker than the Flory number $P_0$ of overlapping molecules at preparation conditions. Since a polymer cannot entangle with more molecules as it overlaps, $f_n$ must converge towards $P_0$ for large N. However, the point at which squeezing becomes impossible must not coincide with the point at which essentially all overlapping rings are concatenated, since squeezing can be stopped, if there is a sufficiently large number of concatenations with surrounding rings that fix the positions of two overlapping non-concatenated rings. Convergence is nearly reached for the samples with the largest values of $f_n$. To show this convergence, we added the data for

$P_0 \propto f_n \varphi_0^{0.27} \propto \varphi_0^{0.65} N^{1/2}$ in Figure 6 ignoring the weak extra $\varphi_0$-dependence of $P_0$. Obviously, a change in overlap number upon swelling must lead to a non-affine contribution to swelling. In Figure 7, we display the apparent affine deformation part $\overline{Q}_a/Q$ of swelling given by

$$\overline{Q}_a = (D/D_0)^3, \qquad (5)$$

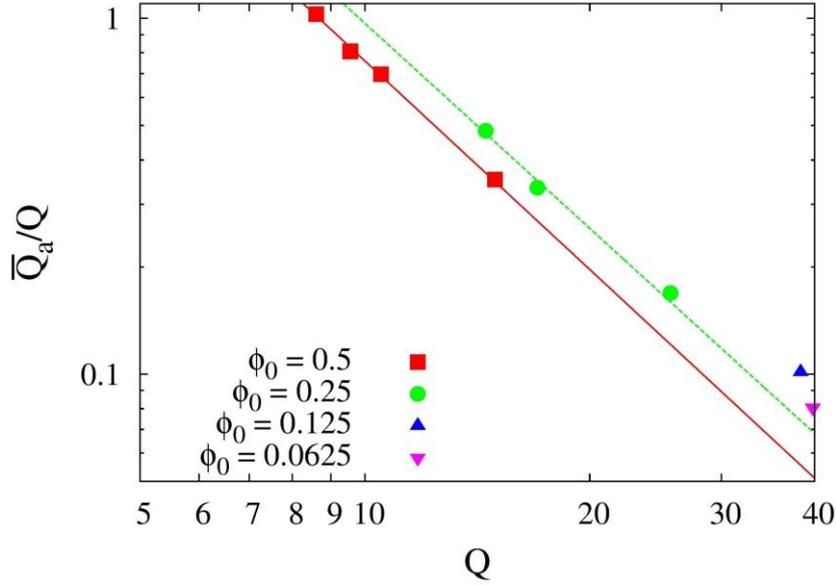

Figure 7. The fraction of the apparent affine contribution $\overline{Q}_a$ to the equilibrium degree of swelling Q. The lines indicate best fits with power laws $\propto Q^{-1.9\pm 0.2}$ and $\propto Q^{-1.95\pm 0.08}$ for $\varphi_0 = 1/4$ and $\varphi_0 = 1/2$ respectively.

Where D and $D_0$ are the ring extension at swelling equilibrium and at preparation conditions respectively. In the FR-model, a fully affine swelling is assumed[31,32] such that $Q \equiv \overline{Q}_a$ and thus, $\overline{Q}_a/Q$ is not a function of Q. The data of Figure 7 indicate a relation in the vicinity of $\overline{Q}_a/Q \propto Q^{-1.95}$ for all samples of our study with a small additional correction as function of $\varphi_0$. Such kind of dependence was not predicted previously and establishes a crucial test for a model explaining the equilibrium swelling of Olympic gels. According to the data of Figure 7 one can expect that for sufficiently long molecules the deformation at equilibrium swelling becomes affine when analyzed on the length scale of the ring diameter. Note that this type of behavior is predicted by some models of rubber elasticity[31,32] (slip link and slip tube models) and that the transition point should be connected to the "affine length" of these models.

The above observations are in agreement with a des-interspersion process originally proposed by Bastide[33] and confirmed by many other studies[34-36] just to name a few. Such a des-interspersion process allows polymer rings to swell in part at no elastic deformation. We argue that the dominating contribution to the non-affinity stems from the des-interspersion of non-concatenated rings upon swelling in the partially concatenated regime where $f_n \propto N$. To derive the equilibrium swelling condition in this regime, we assume full des-interspersion of overlapping non-concatenated rings while concatenated rings remain in overlap. The des-interspersion process may occur in parallel with (or is followed by) an affine deformation of the concatenated rings. For the sake of the argument let us separate des-interspersion from deformation and introduce an intermediate state of swelling that we call the "des-interspersed state". This state is denoted by a subscript 'des' in the following. In the des-interspersed state, the total number of correlation volumes per volume of a ring, $R_{des}^2/\xi_{des}^3$, can be approximated by the number of blobs per chain, $N/g_{des}$ times the number $f_n$ of overlapping concatenated chains

$$\left(\frac{R_{des}}{\xi_{des}}\right)^3 \approx \frac{N}{g_{des}} f_n. \qquad (6)$$

This equation defines the polymer volume fraction at the des-interspersed state, $\varphi_{des}$, that leads[16] to a degree of swelling in the des-interspersed state

$$Q_{des} = 1/\varphi_{des} \propto N^{-(3\nu-1)}\varphi_0^{-2\nu} \propto N^{-0.76}\varphi_0^{-1.18}. \qquad (9)$$

This result is the key to understand the negative power for N at the equilibrium degree of swelling.

For $f_n > 1$, des-interspersion must stop at a polymer volume fraction larger than the overlap concentration $\varphi^* \propto N^{-(3\nu-1)}$ proposed by De Gennes [3]. In fact, we find $\varphi_{des} \propto 1/\varphi^*$, which shows that des-interspersion becomes increasingly difficult with increasing overlap of the rings. Since $\varphi_{des} \gg \varphi^*$, swelling equilibrium can only be reached by an additional elastic deformation of the rings.

To model deformation of the rings, we consider only the permanent entanglements as approximated by the number of concatenations to be relevant at swelling equilibrium and assume that higher topological invariants are not important for the partially concatenated regime with $f_n \propto N$. To apply the affine model for deformation, we subdivide the N segments of the ring into $f_n$ elastic chains by assuming that for small $f_n \leq 10$ all concatenating rings are deforming the concatenated ring at swelling equilibrium. Swelling equilibrium is found by using $\varphi_{des}$ as new "preparation condition" instead of $\varphi_0$ in Eq. (1). This leads to[16]

$$Q \approx \left(\frac{N}{f_n(\varphi_0)}\right)^{3(3\nu-1)/4} \varphi_{des}^{-1/4} \approx N^{-(3\nu-1)/4}\varphi_0^{-5\nu/4} \approx N^{-0.19}\varphi_0^{-0.74}, \qquad (10)$$

which is in good agreement with our ad hoc scaling prediction in Eq.(4) for the simulation data. As direct consequence of this model we find that the apparent affine fraction of swelling depends on the des-interspersed state

$$\overline{Q}_a = \left(\frac{R_g}{R_{g,0}}\right)^3 = \left(\frac{R_{des}}{R_{g,0}}\right)^3 \left(\frac{R_g}{R_{des}}\right)^3 = \left(\frac{R_{des}}{R_{g,0}}\right)^3 Q_a. \tag{11}$$

Since the true affine fraction $Q_a/Q = 1/Q_{des}$ is related to the equilibrium degree of swelling by $Q_a(Q)/Q \propto Q_{des}^{-1}(Q) \propto Q^{-4}$, see Eq.(10), the apparent affine fraction of swelling becomes a universal function of Q:

$$\frac{\overline{Q}_a}{Q} = \left(\frac{R_{des}}{R_{g,0}}\right)^3 \frac{Q_a}{Q} \propto \varphi_0^{3(\nu-1/2)/(3\nu-1)} Q^{-2/(3\nu-1)}. \tag{12}$$

with a strong dependence, $Q_a/Q \propto Q^{-2.62}$, on Q. A similar but weaker dependence on Q was observed in Figure 7, which is a striking evidence for the impact of a des-interspersion process as proposed above on the swelling equilibrium of Olympic gels.

In summary of the above observations we conclude that Olympic gels can display a highly non-affine swelling behavior due to a des-interspersion processes, if the linking number $f_n$ remains below a critical number at which overlapping, mutually not concatenated rings become fixed in space by the concatenations with the surrounding rings. The acceptable agreement between simulation data and our simple model further indicates that each concatenation may contribute a pair of elastic strands to the network, which might be a reasonable approximation for the partially concatenated regime. Improved estimates for the equilibrium degree of swelling of Olympic gels could be obtained by using a more advanced description of the deformations of the cyclic molecules including, for instance, the slippage of entanglements as modeled in slip link or slip tube models of rubber elasticity[31,32].

We expect that our equilibrium swelling data help to improve existing models of rubber elasticity. This expectation stems from the fact, that the structure of any network can be decomposed into a set of connected cycles[17,18] whereby the average cycle size is of the order of 8 chains for typical strand lengths around 50-100 Kuhn segments between 4-functional junctions[19,20]. Therefore, most elastomers are located in the regime $f_n \propto N$ where des-interspersion of non-concatenated cyclic structures upon swelling occurs. Based upon our results, therefore, we expect a clear impact of des-interspersion onto the equilibrium swelling degree of polymer gels. This view is supported by simulations that detect a non-affine swelling of cross-linked networks on length scales much larger than the size of individual network strands[40] and by experiments that measure a vanishing non-affine contribution to elasticity at large degrees of swelling[41,42]. Scattering and NMR data indicate that the initial swelling may be dominated by a des-interspersion process that is followed by a deformation of the chains[33-36], which is in full accord with our model but in opposite order as the assumed by Painter and Shenoy[43].

Another open problem closely related to our work is the explanation of the peak in the swelling activity parameter (dilational modulus) that was addressed in numerous experimental studies as summarized in Refs. [44-46]. In the past, different explanations have been brought forward to explain the nature of the elusive peak. In order to account for the "over-scattering" of gels as compared to solutions[47], Pekarski *et al.*[45] split the osmotic pressure of gels into a "network" and a "liquid" contribution that leads to an effective dependence of the net osmotic pressure on cross-link concentration. However, their arguments apply only for the limiting case of cross-linking at low polymer concentration where the effect of entanglements can be ignored. Xu *et al.*[46] argue that the apparent peak in the swelling activity parameter might be caused by an incorrect data smoothing in previous works. However, the approach of Xu *et al.* contradicts (by assuming a fully affine deformation of the polymers) all experimental and simulation results that determined a non-affine swelling of gels or polymer networks directly or indirectly[33-36,40-43,47].

Experimental data on Olympic gels could be of large importance to settle the discussion about the swelling activity parameter, since a) there are no cross-links that might lead to a different effective interaction parameter (different interactions or larger polymer density near cross-links), b) the degree of non-affine deformations varies largely with Q, and can be estimated using our work, and c) only entanglements are important for the properties of Olympic gels. In contrast to crosslinks connecting two network strands, an entanglement between two strands restricts the motion of the polymers typically in some directions while leaving other directions unconstrained. To understand this particular difference is necessary model the swelling of entangled polymer gels.

Our analysis of Olympic gels gives a fresh view on the problem of swelling of polymer gels in general and reveals that connectivity caused by topological concatenation can lead to a qualitatively different swelling behavior.

## Acknowledgement
ML thanks T. Kreer, A. Galuschko, and M. Gottlieb for stimulating discussions and the DFG for funding grant LA2735/2-1.

[1] E. Raphael, C. Gay, P. G. de Gennes, *J.Stat.Phys*. **89**, 111 (1997).
[2] G. T. Pickett, *Europhys.Lett*. **76**, 616 (2006).
[3] P. G. De Gennes, *Scaling Concepts in Polymer Physics*, Cornell University Press, Ithaca (1979).
[4] M. Rubinstein, J. Pol. Sci B **48**, 2548 (2010).
[5] M. Kapnistos, M. Lang, D. Vlassopoulos, W. Pyckhout-Hintzen, D. Richter, D. Cho, T. Chang, M. Rubinstein, *Nature Materials*, **7**, 997, (2008).
[6] J. E. Mark, *New. J. Chem*. **17**, 703 (1993).
[7] M. Bohn, D. W. Heermann, *J. Chem. Phys.* **132**, 044904 (2010).
[8] M. Bohn, D. W. Heermann, O. Lourenco, C. Cordeiro, *Macromolecules* **43**, 2564 (2010).
[9] A. Narros, A. J. Moreno, C. N. Likos, *Biochem. Soc. Trans.* **41**, 630 (2013).
[10] M. Lang, J. Fischer, J.-U. Sommer, *Macromolecules* **45**, 7642 (2012).
[11] J. Lukes, D. L. Guilbride, J. Votypka, A. Zikova, R. Benne, P. T. Englund, *Eukaryotic Cell*, **1**, 495 (2002).
[12] B. Liu, Y. Liu, S. A. Motyka, E. E. C. Agbo, P. T. Englund, *Trends in Parasitology*, **21**, 363 (2005).
[13] Y. Diao, K. Hinson, R. Kaplan, M. Vasquez, J. Arsuaga, *J. Math. Biol.* **64**, 1087 (2012).
[14] J. Arsuaga, Y. Diao, K. Hinson, *J. Stat. Phys.* **146**, 434 (2012).
[15] Y. Diao, K. Hinson, J. Arsuaga, *J. Phys. A,* **45**, 035004 (2012).
[16] M. Lang, J. Fischer, M. Werner, J.-U. Sommer, *Phys.Rev.Lett*. **112**, 238001 (2014).
[17] S. Nedelcu, M. Werner, M. Lang, J.-U. Sommer, *J.Comp.Phys*. **231**, 2811 (2012).
[18] I. Carmesin, K. Kremer, *Macromolecules* **21**, 2819 (1988).
[19] W. Paul, K. Binder, D. W. Heermann, K. Kremer, *J.Phys*. *II* **1**, 37 (1990).
[20] M. Lang, W. Michalke, S. Kreitmeier, *J. Comp. Phys*. **185**, 549 (2003).
[21] M. Lang, *Macromolecules* **46**, 1158-1166 (2013).
[22] S. J. Clarson, J. E. Mark, J. A. Semlyen, *Polymer Communications* **27**, 244 (1986).
[23] T. J. Fyvie, H. L. Frisch, J. A. Semlyen, S. J. Clarson, J. E. Mark, *J. Polym. Sci. A* **25** 2503 (1987).
[24] L. Liu, P. Li, S. A. Asher, *Nature* **397**, 141 (1998).
[25] H. L. Frisch, E. Wassermann, *J. Am. Chem. Soc.* **83** 3789 (1961).
[26] F. Ferrari, H. Kleinert, I. Lazzizzera, *Phys. Lett. A* **276**, 31 (2000).
[27] F. Ferrari, H. Kleinert, I. Lazzizzera, *Eur. Phys. J. B* **18**, 645 (2000).
[28] M. Otto, *J. Phys. A* **34**, 2539 (2001).
[29] P. J. Flory, J. Rehner Jr., *J.Chem.Phys*. **11**, 521 (1943).
[30] M. Rubinstein, R. Colby, *Polymer Physics*, Oxford University Press, Oxford, UK (2003).
[31] M. Rubinstein, S. Panyukov, *Macromolecules* **30**, 8036 (1997).
[32] M. Rubinstein, S. Panyukov, *Macromolecules* **35**, 6670 (2002).
[33] J. Bastide, C. Picot, S. Candau, *J.Macromol.Sci.Phys*. **B19**, 13 (1981).
[34] J. P. Cohen-Addad, M. Domard, J. Herz, *J.Chem.Phys*. **76**, 2744 (1982).
[35] J. P. Cohen-Addad, M. Domard, G. Lorentz, J. Herz, *J. Physique* **45**, 575 (1984).
[36] W. Chasse, S. Schlögl, G. Riess, K. Saalwächter, *Soft Matter* **9**, 6943 (2013).
[37] M. Lang, W. Michalke, S. Kreitmeier, *Macromol. Theo. Simul*. **10**, 204 (2001).
[38] M. Lang, W. Michalke, S. Kreitmeier, *J. Chem. Phys*. **114**, 7627 (2001).
[39] M. Lang, S. Kreitmeier, D. Göritz, *Rubber Chem. & Tech*. **80**, 873 (2008).
[40] J.-U. Sommer, S. Lay, *Macromolecules* **35**, 9832 (2002).
[41] S. M. Gumbell, L. Mullins, R. S. Rivlin, *Trans. Faraday Soc*. **49**, 1495 (1953).
[42] G. Allen, M. J. Kirkham, J. Padget, C. Price, *Trans. Faraday Soc*. **66** 1278 (1970).
[43] P. C. Painter, S. L. Shenoy, *J. Chem. Phys*. **99**, 1409 (1993).
[44] M. Gottlieb, R. J. Gaylord, *Macromolecules* **17**, 2024 (1984).
[45] P. Pekarski, Y. Rabin, M. Gottlieb, *J. Phys. II* **4**, 1677 (1994).
[46] B. Xu, X. Di, G. B. McKenna, Macromolecules **45**, 2402 (2012).
[47] J. Bastide, S. Candau, *The Physical Properties of Polymer Gels*, Chapter 5, J. Wiley & Sons, New York (1996).